# Combining features of the Unreal and Unity Game Engines to hone development skills

Ioannis Pachoulakis and Georgios Pontikakis

Department of Informatics Engineering
Technological Educational Institute of Crete
Heraklion, Crete 71500, Greece
*{ip@ie.teicrete.gr, pontikakis.gewrgios@gmail.com}*

**Abstract:** Two of the most popular game engines today, Unreal Engine v4.x and Unity Game Engine v5.x have recently adopted competitive and very appealing pricing structures for individual game developers and small teams. One may lean towards one or the other game engine based on various criteria: existing familiarization / vested interest, steepness of learning curve, quality, richness, variability and pricing of add-on assets, initial cost of basic ownership versus subscribing to updates and acquiring needed modules, etc. In this paper we present complementary features of the Unreal and Unity game engines that can be combined to enhance understanding game development and to hone relevant skills. Such combinations of different game engine features can increase the added value for our undergraduate informatics engineering students taking a "Game Technologies" class.

## 1. INTRODUCTION

Game Technologies is a 6$^{th}$ semester course at the Department of Informatics Engineering of TEI of Crete. Its purpose is to draw on the students' background on object-oriented and event based programming, combine these skills with physics and mathematics and produce game design and implementation. Assuming that the idea of creating a game can seem very compelling to the students, we have identified implementation features of the Unreal and Unity game engines that can be nicely combined to hone algorithm understanding and game development skills.

Unreal 4.x by Epic Games [1] is a very successful game engine (freely downloadable at www.unrealengine.com) used both by indy developers and by large game teams. Activating the free license binds the user to paying a 5% royalty fee on gross revenue for any game or application shipped after the first $3,000 per product, per quarter. Unreal Engine 4.x allows game development using the C++ programming language or a form of visual scripting called Blueprint (reminiscent of an algorithmic flowchart). This type of visual scripting can be used for the development of an entire game or just parts of it. Blueprint's capabilities may still be somewhat limited compared to using C++, but they are sufficient to quickly develop simple games. Also, Unreal's production team seems interested in growing Blueprint.

Unity [2], on the other hand, comes in two versions: personal and professional. Of interest to our engineering students is that the personal edition "may not be used by an individual (not acting on behalf of a Legal Entity) or a Sole Proprietor that has reached annual gross revenues in excess of US$100,000 from its use of the Software during the most recently completed fiscal year, which does not include any income earned by that individual which is unrelated to its use of the Software". Unity uses mostly C# or JavaScript, where our students are sufficiently apt. Which game engine offers the shortest path to a finished game depends a lot on personal preference and experience. Both C++ (Unreal) and C# (Unity) are powerful and optimized object oriented programming languages. In this paper we compare a simple pool game implemented in Unreal Engine (using Blueprint) and in Unity (using C#) and show how

one can arrive at functionally equivalent results via visual scripting and event based programming. Our intent is to construct a collection of mini-games exposing different aspects of such functional equivalence to the benefit of our Game Technologies students.

**2. THE POOL GAME**

For the needs of the present article, we adopted the demo game built in Unity's first official beginner tutorial, called "Roll-a-Ball" (https://unity3d.com/learn/tutorials/modules) and recreated that same game in Unreal, this time using Blueprint. The game begins with a stationary white cue ball centered on a pool table. Twelve cubes are placed in a circular pattern around the cue ball floating slightly above the pool table and spinning around their centers. The objective of the game is simple: use the arrow keys to move the cue ball so as to hit and destroy all cubes. Sample snapshots of the game appear in Figure 1.

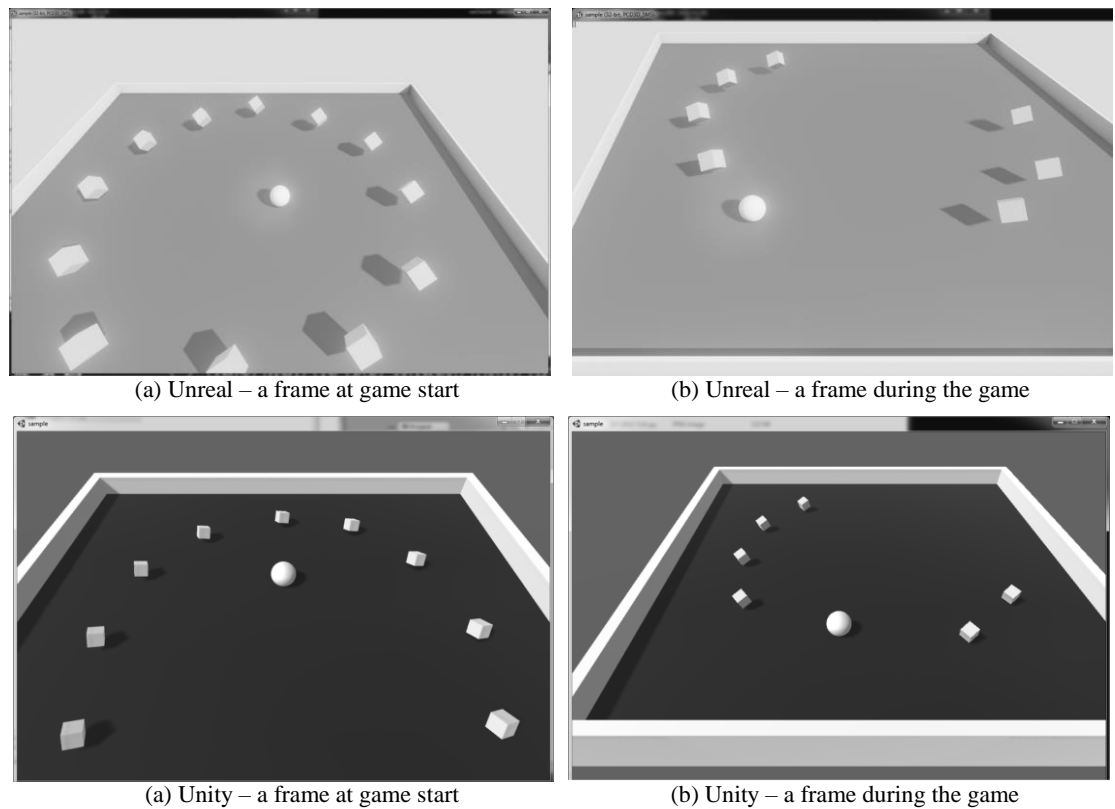

(a) Unreal – a frame at game start     (b) Unreal – a frame during the game

(a) Unity – a frame at game start     (b) Unity – a frame during the game

**Figure 1: The top/bottom panels show sample frames of the Unreal/Unity version of the game. The left panels show a frame at game start and the right panels a sample frame in the duration of the game**

The following three key game functionalities implemented in both (Unreal and Unity) game environments are discussed below:
- Spinning the cubes around their centers
- Rolling of the cue ball on the pool table
- Removing a cube removal upon after a collision with the cue ball

*2.1. Cube Spinning*

Unreal's Blueprint implementing cube spin appears in the top panel of Figure 2. In that panel, the "Event Tick" is triggered in every frame and returns the time interval between the current frame and the previous one. The resulting value, "Delta Seconds" and the value of the "Rotation Factor" are fed via paths 2 and 3, respectively, to the X multiplier. The result of the multiplication is then fed via path 4 to the Yaw value of a vector3 creator used for rotation vectors. The resultant vector3 is applied to the object (parent cube) using the function "AddWorldRotation" as a parameter via path 5. Path 1 defines for the execution sequence.

For the Unity game version, the bottom panel of Figure 2 shows how cube spinning is implemented using a Rotator attached to the cube. The Update() function of the Rotator is called once per frame (before rendering the frame) and performs a rotation transform using three Euler angles. Multiplication of the rotation vector by Time.deltaTime makes the action smooth and frame-independent.

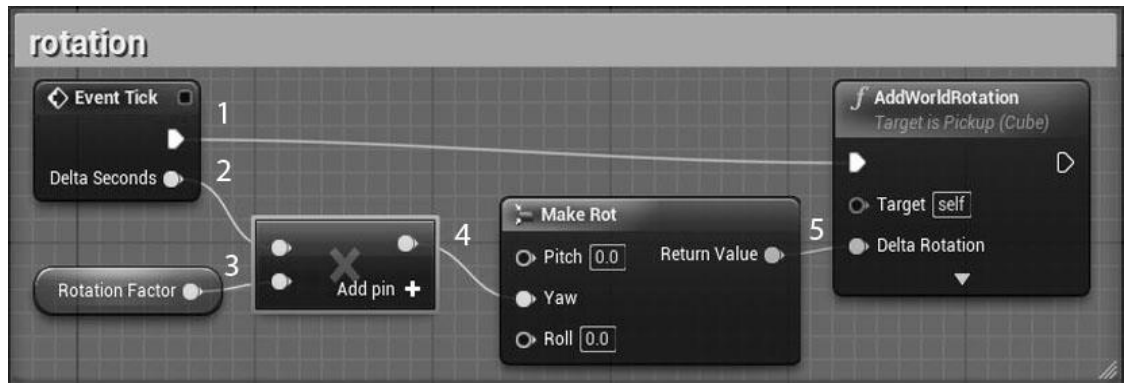

```
using UnityEngine;
using System.Collections;
public class rotator : MonoBehaviour {
  void Update()  {
    transform.Rotate(new Vector3(15, 30, 45) * Time.deltaTime);
  }
}
```

**Figure 2: Constructs implementing a cube's rotation around its center using Unreal's Blueprint (top) and a C# script in Unity (bottom)**

*2.2. Cue ball Roll*

Unreal's Blueprint implementing cue ball roll appears in the top two panels of Figure 3. In the top panel labeled "Roll left/right", the event "InputAxis MoveRight" is triggered by pressing the Left/Right arrow key and returns the value 1 or -1, respectively. This value is fed to the "X multiplier" along with the "RollTorque" value via paths 2 and 3 respectively. The result of the multiplier is fed via path 4 to the X-axis value of the vector3 creator called "Make Vector" (the Y and Z values are left to zero). This vector is then passed via path 6 to the "Add Torque" function along with the Ball object via path 5. The function then applies the torque vector to the ball. Path 1 connects the event with the function and it drives the execution of the algorithm.

The middle panel labeled "Roll forwards/backwards" works in a similar fashion, the only difference being that the "InputAxis MoveForward" event is now triggered by pressing the Up/Down arrow key (forward=1 backward=-1), which is now fed to the Y-axis value of the "Make Vector" vector3 creator.

Equivalent functionality for the Unity version of the game appears in the FixedUpdate() function of the C# script attached to the cue ball (bottom panel of Figure 3). Whereas the Update() function we saw in the case of spinning the cube is called once per frame to update drawable objects, FixedUpdate() is called when we want to employ physics (e.g., forces, collisions, etc.). First, however, a Rigidbody component is attached to the cue ball game object to allow it to respond to physics (this is done at design level and is not visible in the code below). The Start() function makes variable rb to refer to the Rigidbody component of the cue ball game object. Having done that, the first two lines in FixedUpdate() read the values corresponding to which (if any) arrow keys have been pressed and pass them to the x and z arguments of a vector3 object named "movement" (Unity uses a y-up world). Finally, having the direction along which to apply force in "movement", we multiply by a "speed" value to fine tune the response of the cue ball in actual game play and apply the result as a force to the Rigidbody component of the cue ball game object. Unity's game engine then takes over to calculate the motion of the ball based on the force applied and the ball's current position and speed.

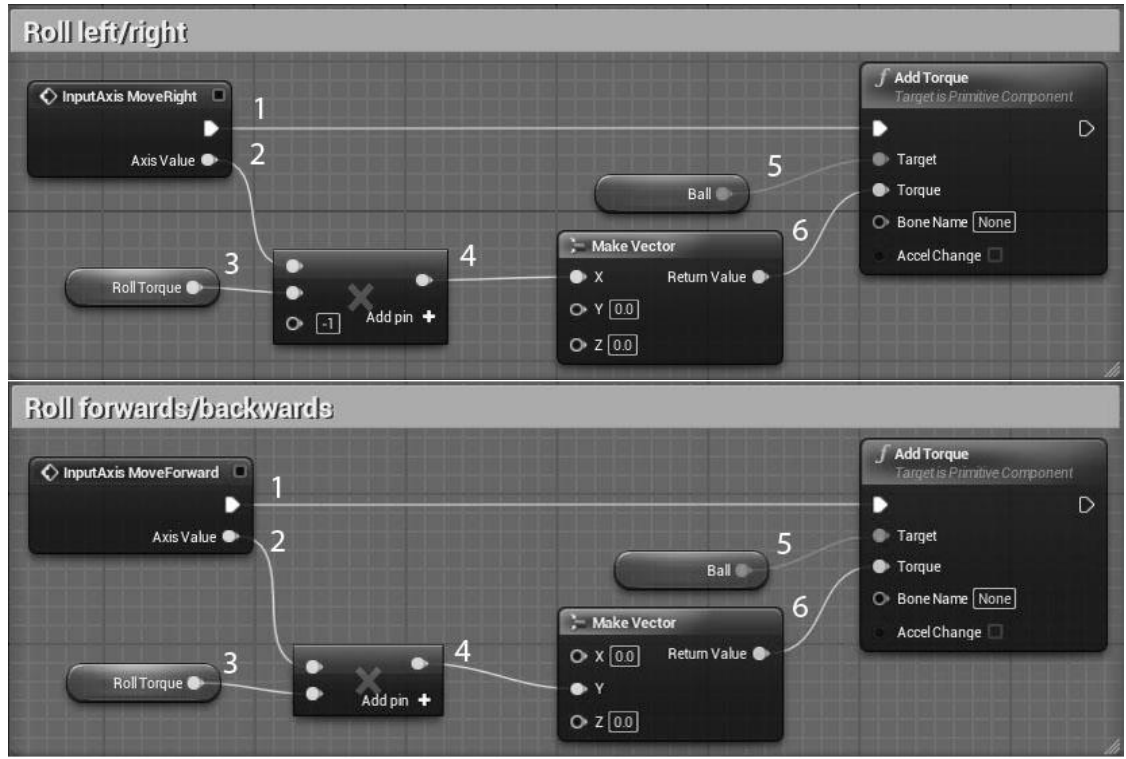

```
using UnityEngine;
using System.Collections;
public class playercontroller : MonoBehaviour {
  public float speed;
  private Rigidbody rb;
  void Start() {
     rb = GetComponent<Rigidbody>();
  }
  void FixedUpdate() {
     float moveHorizontal = Input.GetAxis("Horizontal");
     float moveVertical = Input.GetAxis("Vertical");
     Vector3 movement = new Vector3(moveHorizontal, 0.0f, moveVertical);
     rb.AddForce(movement * speed);
  }
  void OnTriggerEnter(Collider other)  {
     if (other.gameObject.CompareTag ("Pick Up"))     {
        other.gameObject.SetActive (false);
     }
  }
}
```

**Figure 3: Top two panels: Blueprint flows to roll the ball in two directions – Bottom panel: How to achieve the same thing in Unity using FixedUpdate()**

*2.3. Cube Removal*

Unreal Engine's Blueprint implementing cube removal when hit by the cue ball appears in Figure 4. The event "ActorBeginOverlap" is triggered when the parent object (cube) overlaps with another object (it could be any other object, but since the cue ball game object is the only one that moves on the table it practically is triggered only by the cue ball). Following the trigger, the parent game object (cube) is destroyed.

Equivalent functionality for the Unity version of the game appears in the OnTriggerEnter() function of the cue ball's SphereCollider component. The code for OnTriggerEnter() is the last function in the C# script attached to the cue ball (bottom panel of Figure 3). Its argument is another collider (the collider component of the object the cue ball collided with) which we obtain access to via the variable "other". If the "other" game object is a cube, we want to inactivate it. The reason we must check is that the cue ball may also collide with the rails of the pool table. How to do this is simple: all twelve cubes are made out of a prefab cube having a tag "Pick Up". Therefore, if the game object corresponding to the collider "other" has a "Pick Up" tag, the last code line in the OnTriggerEnter() function de-activates that (cube) object.

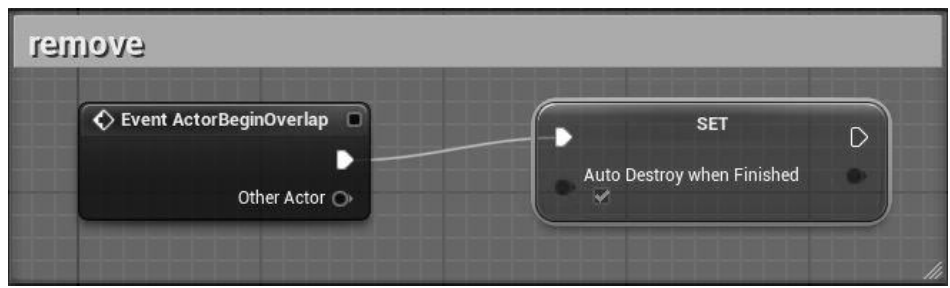

**Figure 4: Blueprint flow for cube removal**

### 3. DISCUSSION AND FUTURE PLANS

Both the Unreal and Unity game engines are full-fledged products which support game development at any level. Unity is generally seen as the more intuitive and easier to grasp game engine, however, Unreal Engine 4's complete UI overhaul has brought with it a very easy to understand UI.

In this paper we have explored the idea of functionality equivalence of tasks required in the development of a simple game in Unreal Engine (using Blueprint visual scripting) and in Unity (using C#). We intend to develop a small collection of mini-games exposing different aspects of such functional equivalence to be used in the "Game Technologies" course taught at the Department of Informatics Engineering at TEI of Crete.